\documentclass[
 reprint,
superscriptaddress,
 amsmath,amssymb,
 aps,
 prl,
]{revtex4-2}

\usepackage{graphicx}
\usepackage{dcolumn}
\usepackage{xcolor}
\usepackage{bm}
\usepackage{siunitx}
\usepackage{amssymb}
\usepackage{amsmath}
\usepackage{hyperref}
\usepackage{url}

\usepackage{lineno}

\begin{document}

\preprint{APS/123-QED}

\title{Laboratory observation of collective beam–plasma instabilities in a relativistic pair jet}
\author{J. W. D. Halliday}
    \address{Department of Physics, University of Oxford, Parks Road, OX1 3PU, UK}
    \address{Central Laser Facility, STFC Rutherford Appleton Laboratory, Didcot OX11 0QX, UK}
    \email{jack.halliday@stfc.ac.uk}
\author{C. D. Arrowsmith}
    \address{Department of Physics, University of Oxford, Parks Road, OX1 3PU, UK}
    \address{Laboratory for Laser Energetics, Rochester, NY, 14623, USA}
\author{A. M. Goillot}
    \address{European Organization for Nuclear Research (CERN), CH-1211 Geneva 23, Switzerland}
\author{P. J. Bilbao}
    \address{GoLP/Instituto de Plasmas e Fus\~ao Nuclear, Instituto Superior T\'ecnico}
\author{P. Simon}
    \address{GSI Helmholtzzentrum f\"ur Schwerionenforschung GmbH, Planckstra\ss e 1 64291 Darmstadt, Germany}
\author{V. Stergiou}
    \address{Department of Physics, University of Oxford, Parks Road, OX1 3PU, UK}
    \address{European Organization for Nuclear Research (CERN), CH-1211 Geneva 23, Switzerland}
\author{S. Zhang}
    \address{Department of Physics, University of Oxford, Parks Road, OX1 3PU, UK}
\author{P. Alexaki}
    \address{Department of Physics, National and Kapodistrian University of Athens, Panepistimiopolis Zografou, GR-15771, Greece}
    \address{European Organization for Nuclear Research (CERN), CH-1211 Geneva 23, Switzerland}
\author{M. Bochmann}
    \address{BoS GmbH / OuSoCo, M\"orbisch am See, 7072, Austria}
\author{A. F. A. Bott}
    \address{Department of Physics, University of Oxford, Parks Road, OX1 3PU, UK}
\author{S. Burger}
    \address{European Organization for Nuclear Research (CERN), CH-1211 Geneva 23, Switzerland}
\author{H. Chen}
    \address{Lawrence Livermore National Laboratory, 7000 East Ave, Livermore, California, 94550, USA}
\author{F. D. Cruz}
    \address{GoLP/Instituto de Plasmas e Fus\~ao Nuclear, Instituto Superior T\'ecnico}
\author{T. Davenne}
    \address{STFC, Rutherford Appleton Laboratory, Didcot OX11 0QX, UK}
\author{A. Dyson}
    \address{Department of Physics, University of Oxford, Parks Road, OX1 3PU, UK}
\author{A. Ebn Rahmoun}
    \address{European Organization for Nuclear Research (CERN), CH-1211 Geneva 23, Switzerland}
\author{I. Efthymiopoulos}
    \address{European Organization for Nuclear Research (CERN), CH-1211 Geneva 23, Switzerland}
\author{D. H. Froula}
    \address{Rochester Laboratory for Laser Energetics, Rochester, NY, 14623, USA}
\author{J. T. Gudmundsson}
    \address{Science Institute, University of Iceland, Dunhaga 3, IS-107, Reykjavik, Iceland}
    \address{Department of Electromagnetics and Plasma Physics, School of Electrical Engineering and Computer Science, KTH Royal Institute of Technology, SE-100 44, Stockholm, Sweden}
\author{D. Haberberger}
    \address{Rochester Laboratory for Laser Energetics, Rochester, NY, 14623, USA}
\author{T. Hodge}
    \address{AWE, Aldermaston, Reading, Berkshire, RG7 4PR, UK}
\author{S. Iaquinta}
    \address{Department of Physics, University of Oxford, Parks Road, OX1 3PU, UK}
\author{E. E. Los}
    \address{Department of Physics, University of Oxford, Parks Road, OX1 3PU, UK}
    \address{Blackett Laboratory, Imperial College London, SW7 2AZ, United Kingdom}
\author{G. Marshall}
    \address{AWE, Aldermaston, Reading, Berkshire, RG7 4PR, UK}
\author{F. Miniati}
    \address{Department of Physics, University of Oxford, Parks Road, OX1 3PU, UK}
\author{S. Parker}
    \address{Blackett Laboratory, Imperial College London, SW7 2AZ, United Kingdom}
\author{B. Reville}
    \address{Max-Planck-Institut f\"ur Kernphysik, Saupfercheckweg 1, D-69117, Heidelberg, Germany}
\author{P. Rousiadou}
    \address{University of Ioannina, Ioannina, 451 10, Greece}
\author{S. Sarkar}
    \address{Department of Physics, University of Oxford, Parks Road, OX1 3PU, UK}
\author{A. A. Schekochihin}
    \address{Department of Physics, University of Oxford, Parks Road, OX1 3PU, UK}
\author{K. G. Schlesinger}
    \address{BoS GmbH / OuSoCo, M\"orbisch am See, 7072, Austria}
\author{L. O. Silva}
    \address{GoLP/Instituto de Plasmas e Fus\~ao Nuclear, Instituto Superior T\'ecnico}
\author{T. Silva}
    \address{GoLP/Instituto de Plasmas e Fus\~ao Nuclear, Instituto Superior T\'ecnico}
\author{R. Simpson}
    \address{Lawrence Livermore National Laboratory, 7000 East Ave, Livermore, California, 94550, USA}
\author{E. Soria}
    \address{European Organization for Nuclear Research (CERN), CH-1211 Geneva 23, Switzerland}
\author{R. M. G. M. Trines}
    \address{Central Laser Facility, STFC Rutherford Appleton Laboratory, Didcot OX11 0QX, UK}
\author{T. Vieu}
    \address{Max-Planck-Institut f\"ur Kernphysik, Saupfercheckweg 1, D-69117, Heidelberg, Germany}
\author{N. Charitonidis}
    \address{European Organization for Nuclear Research (CERN), CH-1211 Geneva 23, Switzerland}
\author{R. Bingham}
    \address{Central Laser Facility, STFC Rutherford Appleton Laboratory, Didcot OX11 0QX, UK}
    \address{Department of Physics, University of Strathclyde, Glasgow, G4 0NG, UK}
\author{G. Gregori}
    \address{Department of Physics, University of Oxford, Parks Road, OX1 3PU, UK}
   
 \date{\today}
 \begin{abstract}
     We report on a measurement of collective behavior in a relativistic electron–positron pair plasma produced in the laboratory. Using the Fireball platform at CERN’s HiRadMat facility, \SI{440}{\giga\electronvolt} protons were used to generate a ultra-relativistic, charge-neutral, electron positron pair beam which propagated through an ambient RF discharge plasma. Magnetic-field amplification due to a plasma-beam instability was diagnosed using a high-sensitivity Faraday-rotation probe, supported by detailed characterization of the diagnostic impulse response. The measured path-integrated magnetic field agrees quantitatively with predictions from particle-in-cell simulations. The results provide a critical benchmark for models of relativistic beam–plasma interactions in astrophysical contexts such as blazar jets and pulsar-wind nebulae.
 \end{abstract}
\maketitle
\emph{Introduction.} 
Electron-positron pair plasmas are expected to be present in a variety of high-energy astrophysical environments, including in the vicinity of black holes and neutron stars \cite{Bambi:2016lkv, arons1979, weidenspointner2008,Meszaros2002}. Collective pair-plasma processes  are frequently invoked to explain astrophysical phenomena including particle acceleration, magnetic field generation, and to account for the observed non-thermal radiation.

Producing densities of pairs sufficient to access plasma effects in terrestrial settings is challenging. Techniques for high-yield positron production employ a range of drivers, including nuclear reactors \cite{hugenschmidt2012}, particle accelerators \cite{bernardini2004, blumer2022}, and high-power lasers \cite{Chen2015, liang2015, sarri2015, xu2016, peebles2021, jiang2021}. A comprehensive overview of progress in laser-driven methods is given by Chen and Fiuza ~\cite{chen2023}.

\begin{figure}
    \centering
    \includegraphics[]{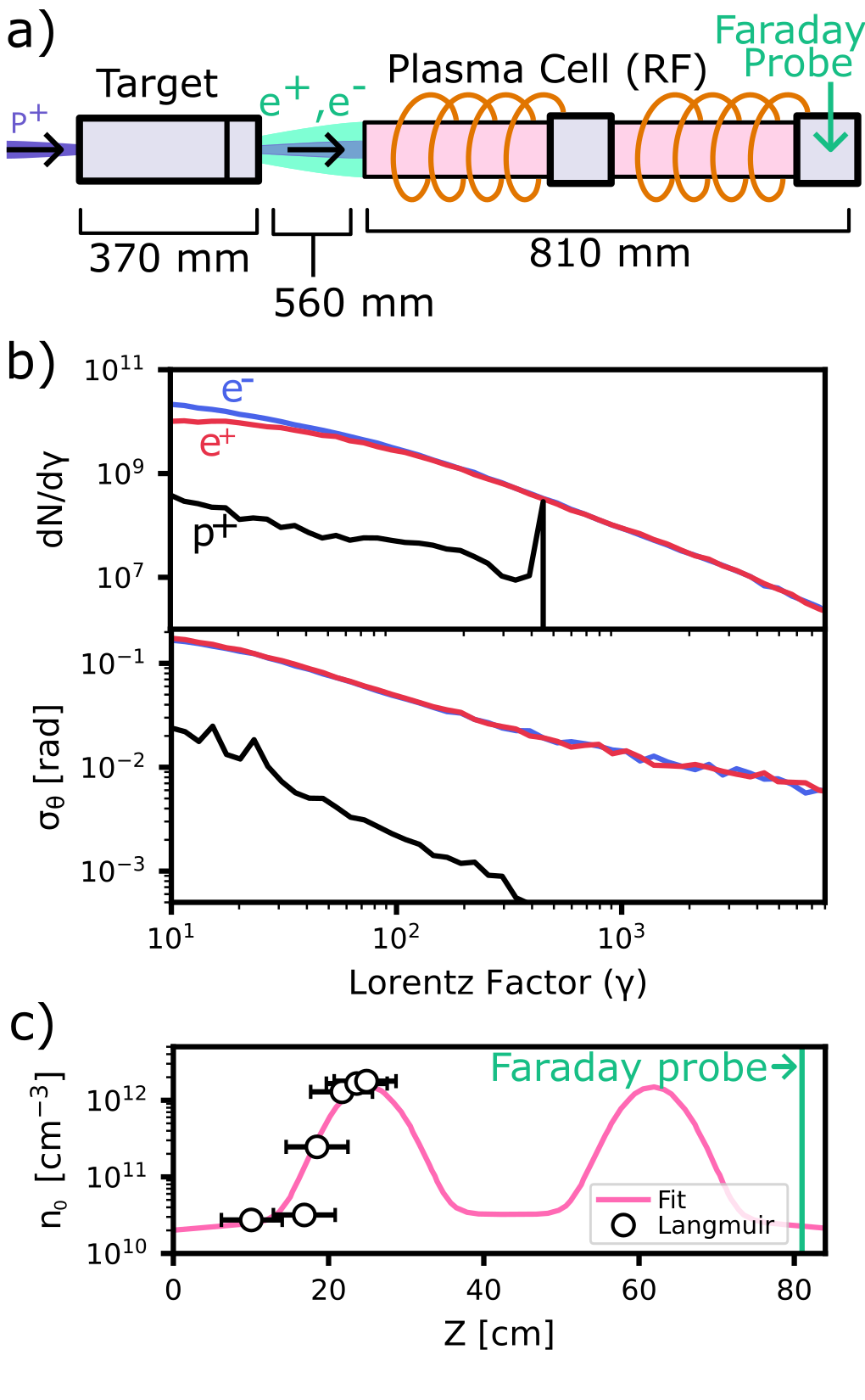}
    \caption{
        \textbf{(a)} Diagram of the Fireball experimental setup, showing the pair-generation target, inductive-mode RF plasma discharge, and Faraday-rotation diagnostic probe. The system was driven by the \SI{440}{\giga\electronvolt} proton beam from the Super Proton Synchrotron at HiRadMat (CERN), delivering a peak intensity of $3\times 10^{11}$ protons per pulse. The beam had a Gaussian transverse profile with \SI{1}{\milli\meter} standard deviation and a Gaussian temporal profile of approximately \SI{250}{\pico\second} duration. \textbf{(b)} Spectra / average divergence for the residual primaries and dominant secondary species produced by the target. These data are from FLUKA (Monte Carlo) simulations and were validated against the experimental measurements reported in Ref.~\cite{Arrowsmith2024}. \textbf{(c)} Profile of electron density in the RF discharge (ambient) plasma, based on Langmuir-probe measurements reported in Ref.~\cite{Arrowsmith2023}.
    }
    \label{fig:setup}
\end{figure}
In this article, we build on previous work \cite{Arrowsmith2024, Arrowsmith2025} using the novel, accelerator based `Fireball' platform for laboratory studies of relativistic pair plasmas. By improving the Fireball diagnostic suite, we were able to measure signatures of magnetic-field amplification due to a beam-plasma instability. To our knowledge, this represents the first measurement of collective behavior in a terrestrial pair-plasma jet.

\emph{Experimental Setup.}
Our experimental setup, depicted in Fig.~\ref{fig:setup}a, relies on ultra‑relativistic protons (momentum \(p = \SI{440}{\giga\electronvolt}/c\)) accelerated in the CERN Super Proton Synchrotron (SPS) and sent to the HiRadMat facility \cite{efthymiopoulos2011}. The figure also shows the inductively coupled radio-frequency (RF) plasma discharge (“plasma cell”), which provides the ambient argon plasma into which the pair beam propagates; its properties will be described later in the text. The proton beam, delivered as a single bunch per extraction, contains \(3\times10^{11}\) particles per pulse with a bunch duration of approximately \SI{250}{\pico\second}, and impinges on a cylindrical target comprising \SI{360}{\milli\metre} of carbon (mass density \SI{1.84}{\gram\per\centi\meter\cubed}) followed by \SI{10}{\milli\metre} of tantalum (mass density \SI{16.7}{\gram\per\centi\meter\cubed}).

The interaction of protons with the target's graphite section initiates a hadronic cascade, including both charged and neutral pions. The longer-lived charged pions ($\pi^\pm$) predominantly undergo further hadronic interactions in the target, whereas neutral pions ($\pi^0$) decay promptly into a directional beam of \si{\giga\electronvolt}-energy $\gamma$-rays. In the tantalum section, these $\gamma$-rays drive an electromagnetic cascade, generating electron-positron pairs. The pair production is further amplified by the Bremsstrahlung of electrons and positrons in the Bethe-Heitler process \cite{Bethe1934}. A more detailed discussion of the optimization process used to design the target is given elsewhere ~\cite{Arrowsmith2021}.

The spectrum and energy-resolved average divergence of the electron-positron pairs, as well as a residual population of primary protons, are shown in Fig.~\ref{fig:setup}b. As expected, the proton spectrum peaks at \SI{440}{\giga\electronvolt}, corresponding to the initial energy of the incident protons. The beam divergence, $\sigma_\theta$, is defined as the standard deviation of $\theta = p_\perp / p_\parallel$, the ratio of transverse to longitudinal momentum components in the laboratory frame. The curves in the plot are based on simulations performed with the Monte Carlo code \textsc{FLUKA}~\cite{bohlen2014, ferrari2005}, and were validated experimentally using data from beam-profile screens and a magnetic spectrometer, as reported in Ref.~\cite{Arrowsmith2024}. The data indicate that electron and positron yields are approximately equal, with pairs having an average Lorentz factor \(\langle \gamma \rangle \sim 500\) and an average divergence \(\langle \sigma_{\theta} \rangle \sim \SI{25}{\milli\radian}\). The energy spectra follow a power-law distribution, the details of which are given in Ref.~\cite{Arrowsmith2024}. 

The total positron yield was \( N_+ = 2.5 \times 10^{13} \), with a positron-to-electron ratio \( N_+ / N_- = 0.8 \), resulting in a peak pair density of \( n_0 = 5 \times 10^{11} \; \si{\per\centi\meter\cubed} \) at the entrance to the plasma cell. This estimate is based on FLUKA-derived transverse fluence profiles, assumes a pair density equal to twice the local positron density (neglecting the small electron excess), and takes the longitudinal profile of the secondaries to match that of the primary protons.

The electron number density within the RF discharge plasma cell (\(n_0\)) as a function of position is shown in Fig.~\ref{fig:setup}c. Details of the cell design and plasma density/temperature measurements are provided in Ref.~\cite{Arrowsmith2023}, where characteristic electron temperatures are reported to be in the range of \(1\)--\SI{10}{\electronvolt}, depending on the operating conditions.

The propagation of the pair plasma through the ambient electron-ion plasma in the experimental cell enables studies of beam–plasma instabilities. Initial experiments described in Ref.~\cite{Arrowsmith2025} found no evidence for instability growth. From this null result, an upper limit of \SI{0.7}{\per\nano\second} was inferred for the growth rate of unstable modes.    
OSIRIS \cite{fonseca2002} particle-in-cell (PIC) simulations were consistent with this bound once the pair-beam divergence -- known to suppress growth \cite{silva2002} -- was included. 

Ref.~\cite{Arrowsmith2025} further outlined a scaling argument relating the Fireball experiment to TeV blazar jets. Even under optimistic assumptions, this analysis shows that the instability growth rate in blazar jets is too small to account for the absence of reprocessed GeV $\gamma$-ray emission \cite{aharonian2006}. This finding contradicts earlier suggestions that beam–plasma instabilities could efficiently dissipate the ultrarelativistic $e^{+}e^{-}$ pairs generated when TeV $\gamma$-rays from blazars interact with the extragalactic background light, thereby suppressing secondary GeV emission \cite{broderick2012, schlickeiser2012}. Consequently, the lower limit on the intergalactic magnetic field inferred from the absence of this GeV component \cite{neronov2010} remains robust. The direct laboratory measurement of magnetic fields in a relativistic pair jet reported here provides quantitative information and supports this interpretation. Moreover we show the measured field amplitudes agree quantitatively with the simulations used to inform the scaling argument in Ref.~\cite{Arrowsmith2025}.

\begin{figure}
    \centering
    \includegraphics[]{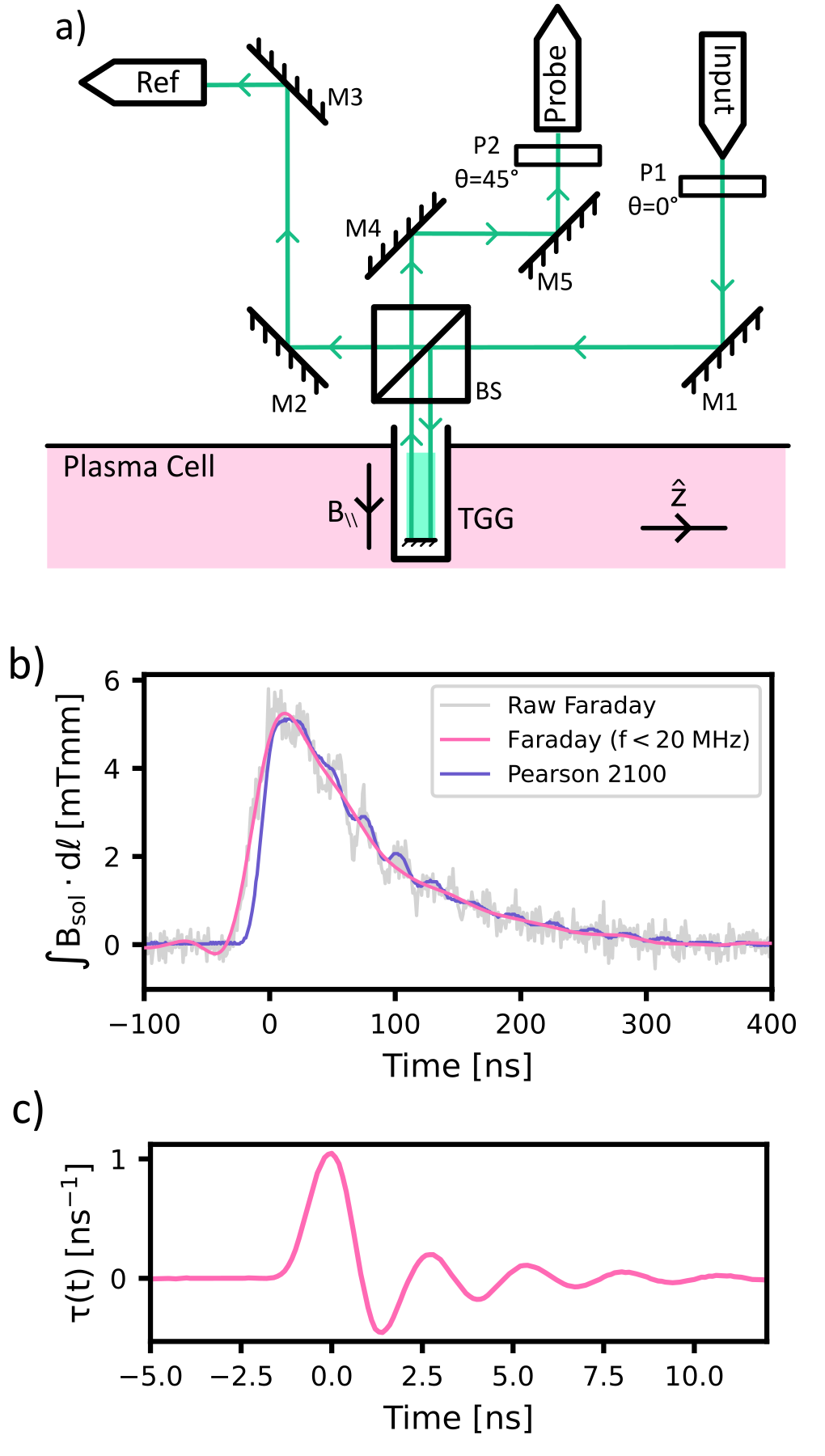}
    \caption{
    \textbf{(a)}~Diagram of the Faraday-rotation diagnostic.
    \textbf{(b)}~Measured Faraday signal in an offline sensitivity characterization 
    \textbf{(c)}~Impulse response function (IRF) of the Thorlabs PDB435A photodioides used in the Faraday setup, normalized so that $\int\tau(t) dt = 1$. 
    }
    \label{fig:faraday}
\end{figure}

\emph{Description of Faraday Rotation Diagnostic.} 
In this work we describe new measurements of magnetic fields in the Fireball setup obtained with a Faraday rotation probe, a setup diagram for which is shown in Fig. \ref{fig:faraday}a. 

A linearly polarized $\lambda = \SI{532}{\nano\meter}$ probe beam is generated by an optically pumped semiconductor laser (COHERENT OBIS) and delivered to the plasma cell via a polarization-maintaining fiber (PANDA PM460-HP). The fiber output is collimated and labeled ``input'' in the figure. The beam passes through a high-efficiency polarizer (Thorlabs LPVISC050, labeled P1), aligned with the fiber's fast axis, and is directed into a cubic beam-splitter (Thorlabs BS010) using protected aluminum mirrors (labeled M1-M5 in the figure). 

The beam splitter transmits one component of the beam to a fiber coupler (labeled ``ref''), relaying it to a detector via a \SI{100}{\micro\meter} core-diameter multimode fiber. The reflected beam propagates through a ceramic reentrant tube immersed in the discharge plasma to a terbium gallium garnet (TGG) cylinder ($L = \SI{12}{\milli\meter}, d = \SI{2}{\milli\meter}$). The TGG is anti-reflection coated on its outward-facing surface and high-reflection coated on the inward surface. The Faraday effect rotates the polarization of light by the angle
\begin{equation}
\alpha = \mathcal{V} \int \mathbf{B} \cdot d\mathbf{l},
\end{equation}
where $\mathcal{V}$ is the Verdet constant of the TGG; for our crystal at the working wavelength we measured $\mathcal{V} = 217\pm 15 ,\si{\radian\per\tesla\per\meter}$, consistent with Ref.~\cite{slezak2016}. The line element $d\mathbf{l}$ is taken along the probe wavevector, so only the component of $\mathbf{B}$ parallel to the beam contributes, and the high-reflection rear surface gives a double pass through the crystal, yielding an effective path length $2L$.

The reflected beam then passes through a second high-efficiency polarizer (P2), oriented at \SI{45}{\degree} relative to P1. By Malus’ law, the polarization rotation caused by the TGG is converted into an intensity change at P2’s output. This output is coupled into a multimode fiber (labeled ``probe'') and relayed to the detector assembly. 

The detector comprises a pair of balanced, current-amplified photodiodes (Thorlabs PDB435A). The probe laser, having passed through the TGG, is incident on one photodiode, while the reference arm is incident on the other. 

Figure~\ref{fig:faraday}b shows an offline sensitivity characterization of the Faraday probe used in this work. For this test, the alumina reentry tube was surrounded by a \SI{10}{\milli\meter}-diameter solenoid. A \SI{2.5}{\ampere} current pulse with a \SI{20}{\nano\second} rise time was driven through the solenoid using an external generator, and the current was monitored with an inductive current probe (Pearson 2100). The path-integrated magnetic field inside the solenoid was calculated numerically using an implementation of the method described in Ref.~\cite{ortner2022}, as provided by the MagPyLib Python library~\cite{magpylib}. The balanced-photodiode signal was then scaled so that its peak coincided with the peak of the $\int \mathbf{B}\cdot d\mathbf{l}(t)$ inferred from the Pearson probe; in this way, a proportionality constant mapping diode voltage to path-integrated magnetic field was obtained.

Since the duration of the driving proton bunch ($\sigma_\tau \sim \SI{250}{\pico\second}$) was comparable to the bandwidth of the photodiodes, it was necessary to characterize their temporal impulse response function (IRF). This is shown in Fig.~\ref{fig:faraday}c, which presents the normalized output of the photodiodes in response to a \SI{50}{\pico\second} laser pulse. The response was measured under transient conditions with no duty cycle, reflecting the single-shot nature of the experiment. Assuming a linear, time-invariant response, the measured Faraday-rotation signal corresponds to the convolution of this IRF with the true path-integrated magnetic field.

In the Fireball experimental setup, the Faraday diagnostic was positioned \SI{800}{\milli\meter} downstream of the entrance to the plasma cell, as shown in Figs.~\ref{fig:setup}a and \ref{fig:setup}c, with the TGG axis oriented perpendicular to the beam-propagation direction.

\begin{figure}
    \centering
    \includegraphics[]{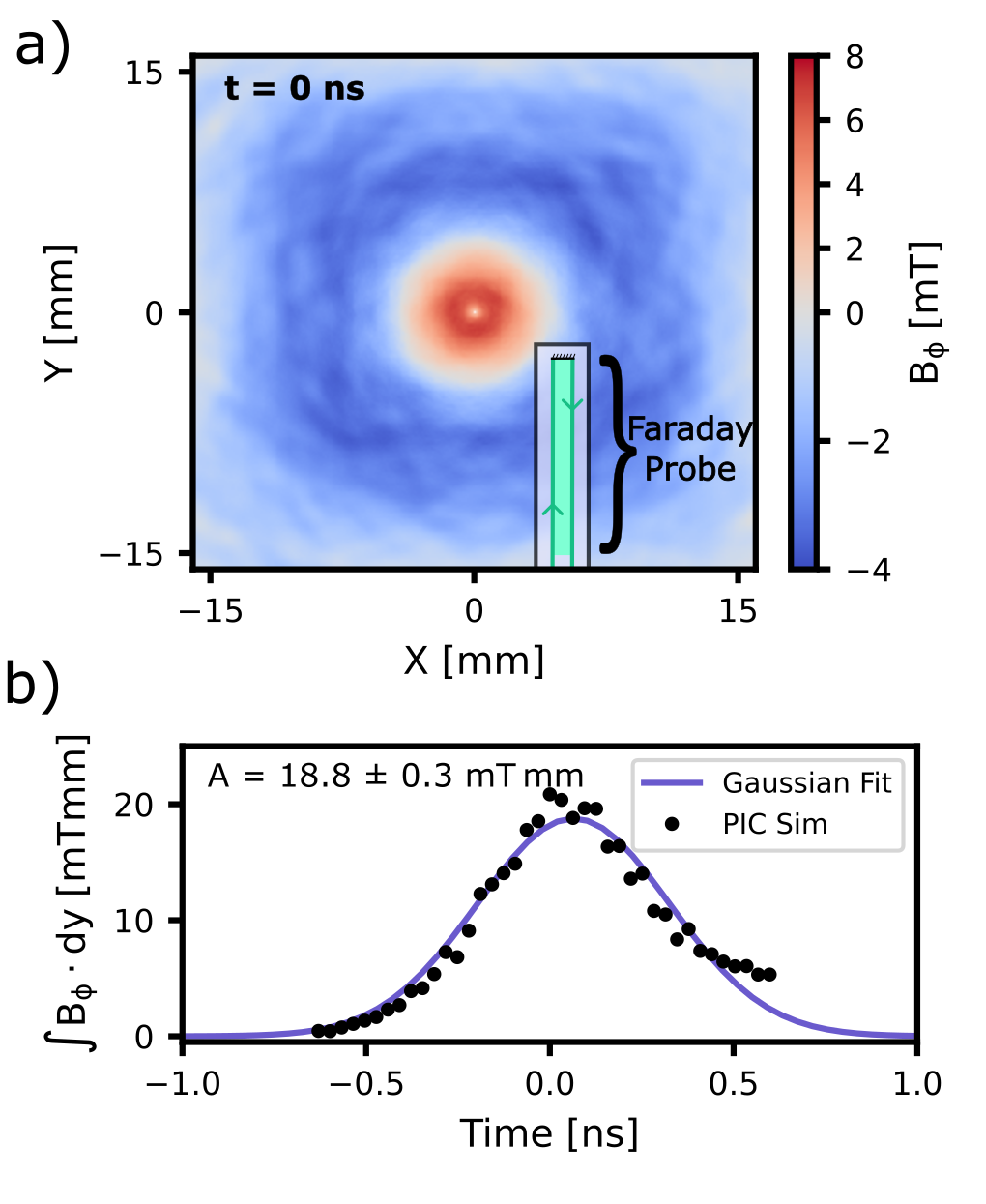}
    \caption{
        Particle-in-cell simulation results. \textbf{(a)} Magnetic field map at the Faraday probe's axial position, with the size and position of the terbium gallium garnet and the alumina re-entrant tube shown to scale. \textbf{(b)} Path-integrated magnetic field at the probe's position calculated from the simulation, fitted to a Gaussian profile.
    }
    \label{fig:B_map_PIC}
\end{figure}

\emph{Results.} 
Three-dimensional particle-in-cell (PIC) simulations were performed with OSIRIS \cite{fonseca2002} in a moving-window geometry propagating at $c$ along $z$, as first described in Ref.~\cite{Arrowsmith2025}. The beam of electrons, positrons, and protons was initialized from analytic fits to FLUKA distributions at the plasma cell entrance, shown in Fig.~\ref{fig:setup}b.  The ambient plasma profile was taken from the Langmuir-probe measurements, shown in Fig.~\ref{fig:setup}c. The simulation domain was $3.5 \times 3.5 \times 40$ cm$^3$ with absorbing boundaries, discretized into $879 \times 879 \times 10050$ cells ($\Delta x = 0.096$ mm), and advanced with $\Delta t = \SI{43.7}{\femto\second}$, satisfying the 3D Courant condition. Each species was represented by 8 particles per cell with quadratic interpolation and first-order current smoothing.

Figure~\ref{fig:B_map_PIC} shows magnetic fields from these simulations. Analysis of the simulations indicates that the azimuthal field grows from an initial seed generated by the on-axis residual proton beam. Figure~\ref{fig:B_map_PIC}a presents a map of the azimuthal magnetic field component in the two dimensions transverse to the beam axis, evaluated at the longitudinal position of the Faraday probe and at the time of peak magnetic field at that location. Overlaid on this field map, to scale, are the dimensions and position of the magneto-optic crystal within the Faraday probe. Experimentally, the crystal’s position was localized with sub-millimeter precision relative to the beam axis using the beamline survey techniques described in Ref.~\cite{gayde2024}.

Figure~\ref{fig:B_map_PIC}b shows the time-resolved magnetic field from the simulations, path-integrated along the axis of the magneto-optic crystal, together with a Gaussian fit to the simulated result. The simulations indicate that the path-integrated fields persist for a time on the order of the driving proton bunch duration, and predict a peak path-integrated field of $\sim \SI{19}{\milli\tesla\milli\meter}$.
\begin{figure}
    \centering
    \includegraphics[]{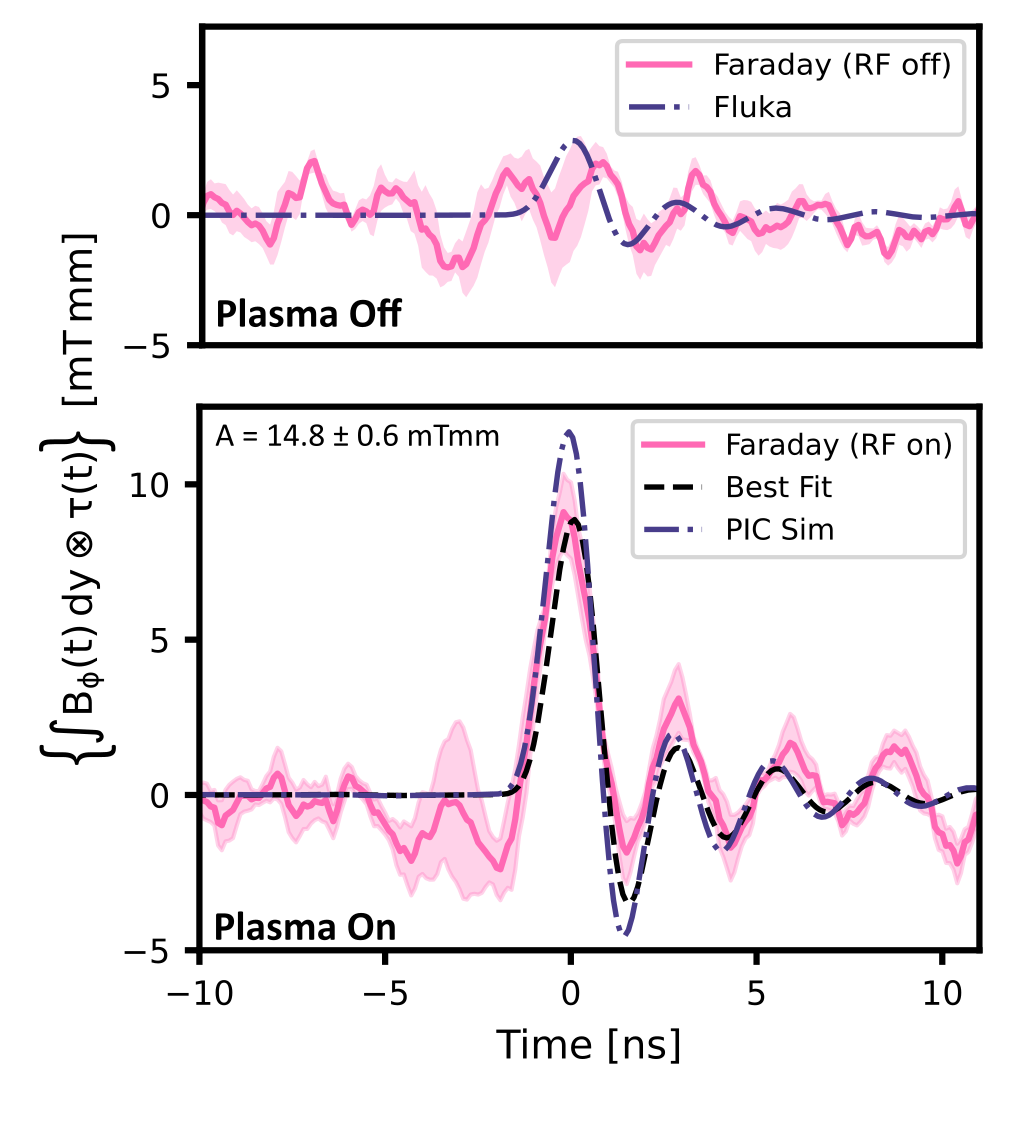}
    \caption{
        Measured Faraday-rotation data for extractions with the plasma on (bottom panel) and with the plasma off (top panel) are shown. The plasma off data are compared to results from FLUKA Monte Carlo simulations, whereas the plasma-on data are compared to results from OSIRIS simulations. For the plasma on case, the best-fit Gaussian, convolved with the impulse response (Fig.~\ref{fig:faraday}c), is also shown. For the experimental data, the shaded band corresponds to the standard deviation across three consecutive repeats. 
    }
    \label{fig:main_data}
\end{figure}

Experimental results obtained via Faraday rotation, for cases with and without the plasma cell energized, are shown in Fig.~\ref{fig:main_data}. These data are constructed by subtracting the average of three shots with the probe laser off from the average of three consecutive shots with the probe laser on; error bars indicate the standard deviation across individual shots, all taken with identical beam conditions. This subtraction removes background fluorescence in the optical fiber induced by the mixed radiation field present during the experiment. The radiation field originates from interactions of the high-energy proton beam with the target and the beam dump downstream of the experimental setup.

In the plasma-off case, no signal above the detector noise floor is observed. These data are compared to a simulated trace derived by applying Ampère’s law to particle fluence maps (electrons, positrons, and protons) obtained from FLUKA Monte Carlo simulations. The model assumes a temporal spread comparable to the \SI{250}{\pico\second} Gaussian profile of the driving proton beam and includes convolution with the measured Faraday probe impulse response (see Fig.~\ref{fig:faraday}c). It thus accounts for free propagation, scattering, and secondary particle production but excludes any collective plasma effects—consistent with the experimental conditions in the absence of ambient plasma. The simulation data agree with the measurements in that no signal is observed above the noise floor.

For the plasma-on data, a clear Faraday-rotation signal is observed. The experimental results are compared to a simulated trace obtained by convolving a Gaussian fit to the simulated, path-integrated magnetic field from the PIC simulations (Fig.~\ref{fig:B_map_PIC}b) with the measured impulse response of the detection system (Fig.~\ref{fig:faraday}c). In this case, the simulation includes the collective interaction between the pair beam and the ambient plasma. 

The fact that magnetic-field growth is observed only when the discharge is active, and is reproduced in simulations only when these collective effects are included, demonstrates that the amplified field is generated by instabilities of the pair beam interacting with the ambient plasma.

Qualitatively, these experimental data agree well with the PIC simulations. To quantify this agreement, we fit the Faraday-rotation data to
\begin{equation}
  f(t) = A \exp\!\left[-\frac{(t - t_0)^2}{2s^2}\right] \otimes \tau(t),
\label{eq:fit}
\end{equation}
where \(A = \int \mathbf{B}_\mathrm{exp}\cdot d\mathbf{l}\) and \(t_0\) are free parameters (with \(t_0\) as the fitted temporal offset accounting for the digitizer delay), \(\otimes\) denotes convolution, and \(\tau(t)\) is the measured IRF (Fig.~\ref{fig:faraday}c). The parameter \(s\) was held fixed at \(\SI{250}{\pico\second}\), the width of the driving proton bunch; the width of the measured signal is dominated by the impulse response, and so allowing \(s\) to vary resulted in poor convergence. With this choice, the width and overall envelope of the fitted signal reflect the temporal evolution of $\int \mathbf{B}\cdot d\mathbf{l}$, while the finer structure in the traces is dominated by the IRF.

The value of the peak amplitude obtained experimentally was $A_{\mathrm{exp}}=14.8\pm \SI{0.6}{\milli\tesla\milli\meter}$. This compares favorably with the prediction from the PIC simulations, $A_{\mathrm{PIC}}=18.8\pm \SI{0.3}{\milli\tesla\milli\meter}$, given that the uncertainties originate purely from the least squares fit and do not account, e.g., for the  systematic error associated with the probe sensitivity, which we conservatively estimate to be on the order of $20\,\%$.

\emph{Discussion \& Conclusions.}
We have presented a direct measurement of collective, plasma-mediated interactions in a relativistic electron–positron pair plasma.

This result builds upon earlier Fireball-platform measurements \cite{Arrowsmith2025}, where no magnetic-field signal was detected and only an upper bound on the instability growth rate could be inferred. In the present experiment, the use of a significantly more sensitive Faraday-rotation probe allows us to estimate the spatially averaged growth rate from the ratio of path-integrated fields:
\begin{equation}
\langle \Gamma_{\mathrm{exp}} \rangle \sim \frac{1}{t_{\mathrm{prop}}}
\ln\!\left[\frac{\int \mathbf{B}_{\mathrm{exp}}\!\cdot d\mathbf{l}}{\int \mathbf{B}_{0}\!\cdot d\mathbf{l}}\right],
\end{equation}
where $\int \mathbf{B}_{\mathrm{exp}}\!\cdot d\mathbf{l}$ is the measured path-integrated field, $\int \mathbf{B}_{0}\!\cdot d\mathbf{l}$ is the corresponding value in the absence of collective effects as obtained from the FLUKA-based free-propagation model described above (Ampère reconstruction of the particle fluence maps), and $t_{\mathrm{prop}}$ is the beam propagation time in the plasma cell. Adopting $t_{\mathrm{prop}}=\SI{2.7}{\nano\second}$, the Gaussian fit in Fig.~\ref{fig:main_data} gives $\int \mathbf{B}_{\mathrm{exp}}\!\cdot d\mathbf{l}=\SI{14.8}{\milli\tesla\,\milli\meter}$ and the FLUKA background (Fig.~\ref{fig:main_data}) gives $\int \mathbf{B}_{0}\!\cdot d\mathbf{l}=\SI{4.9}{\milli\tesla\,\milli\meter}$, yielding $\langle \Gamma_{\mathrm{exp}} \rangle=\SI{0.40}{\per\nano\second}$, in good agreement with PIC simulations, $\langle \Gamma_{\mathrm{PIC}} \rangle=\SI{0.50}{\per\nano\second}$.

For comparison, we use the temporal growth rate of the oblique mode in a homogeneous plasma, including the effect of a finite angular spread \cite{fainberg1970},
\begin{equation}
\Gamma_{\mathrm{L}} = \frac{\sqrt{3}}{2^{4/3}}\,\omega_{\mathrm{p}}\varrho^{1/3}\left[\frac{k_{\perp}^{2/3}}{k^{2/3}}-\frac{3k_{\perp}^2\sigma_\theta^2}{8k_{\parallel}^2}\left(\frac{2}{\varrho}\right)^{2/3}\right],
\label{eq:gamma_l}
\end{equation}
where $\omega_\mathrm{p}$ is the ambient plasma frequency; $\varrho = n_{\pm}/(n_0 \gamma)$ is the pair–to–ambient density ratio evaluated in the pair rest frame; and $\mathbf{k}$ is the perturbation wave vector with magnitude $k\equiv|\mathbf{k}|$. We decompose $\mathbf{k}$ into components perpendicular and parallel to the beam axis, denoted $k_\perp$  and $k_\parallel$ respectively. 
In practice, complications such as the finite beam radius and energy spread further suppress the instability \cite{bret2010}, so Eq.~(\ref{eq:gamma_l}) should be regarded as an upper bound. 

Because the ambient plasma density varies along the cell (Fig.~\ref{fig:setup}c), and the pair density decreases due to emittance, we evaluate the instantaneous growth rate along the plasma and beam-density profiles and define an effective growth rate obtained by averaging the instantaneous rate along the propagation path, 
\begin{equation} 
\Gamma_\mathrm{eff}(t)=\frac{1}{t}\int_0^t \Gamma_\mathrm{L}(t'),\mathrm{d}t', \end{equation} 
formulated such that $B(t)=B_0\exp[\Gamma_\mathrm{eff}(t)\times t]$.

Performing this calculation yields
$\Gamma_\mathrm{eff}(t_\mathrm{prop}) \sim \SI{2.5}{\per\nano\second}$, taking
$k_\parallel = k_\perp = 2\pi/\SI{2}{\milli\meter}$, consistent with the transverse beam size and the diameter of the Faraday probe. The significantly smaller values obtained from PIC simulations and experiment indicate that additional kinetic and geometric effects reduce the true growth rate by a further factor of $\sim 5$, consistent with expectations for relativistic pair-beam instabilities with finite longitudinal spread and finite beam radius.

The agreement between experiment and simulation demonstrates that collective beam–plasma interactions can now be characterised quantitatively in a relativistic pair plasma under laboratory conditions. This capability opens the way to controlled studies of unstable-mode structure, nonlinear evolution, and magnetic-field topology in relativistic pair jets, providing stringent laboratory benchmarks for kinetic models relevant to high-energy astrophysical environments.

\section*{Acknowledgments}
This project has received funding from the European Union’s Horizon Europe Research and Innovation program under Grant Agreement No 101057511 (EURO-LABS). The work of G.G. was partially supported by UKRI under grant no. ST/W000903/1 and EP/Y035038/1, while A.F.A.B. was also supported by UKRI (grant number MR/W006723/1). The work of D.H.F. and D.H. was supported by the U.S. Department of Energy under Award Number DE-NA0004144. The work of P. J. B., F. D. C., L.O.S., and T. S.  was supported by  the FCT (Portugal) under grants I/BD/151559/2021, X-MASER 2022.02230.PTDC, and IPFN-CEEC-INST-LA3/IST-ID. FLUKA simulations were performed using the STFC Scientific Computing Department’s SCARF cluster. OSIRIS simulations were performed at LUMI-C (Finland) within EuroHPC-JU Project No. EHPC-REG-2021R0038, and at Deucalion (Portugal), funded by FCT Masers in Astrophysical Plasmas (MAPs) I.P project 2024.11062.CPCA.A3.
\bibliography{refs}

@article{Arrowsmith2021,
  title = {Generating ultradense pair beams using 400 $\mathrm{GeV}/c$ protons},
  author = {Arrowsmith, C. D. and Shukla, N. and Charitonidis, N. and Boni, R. and Chen, H. and Davenne, T. and Dyson, A. and Froula, D. H. and Gudmundsson, J. T. and Huffman, B. T. and Kadi, Y. and Reville, B. and Richardson, S. and Sarkar, S. and Shaw, J. L. and Silva, L. O. and Simon, P. and Trines, R. M. G. M. and Bingham, R. and Gregori, G.},
  journal = {Phys. Rev. Res.},
  volume = {3},
  issue = {2},
  pages = {023103},
  numpages = {9},
  year = {2021},
  month = {May},
  publisher = {American Physical Society},
  doi = {10.1103/PhysRevResearch.3.023103},
}

@article{Arrowsmith2023,
doi = {10.1088/1748-0221/18/04/P04008},
year = {2023},
publisher = {IOP Publishing},
volume = {18},
pages = {P04008},
author = {C.D. Arrowsmith and A. Dyson and J.T. Gudmundsson and R. Bingham and G. Gregori},
title = {Inductively-coupled plasma discharge for use in high-energy-density science
          experiments},
journal = {Journal of Instrumentation},
}

@article{Arrowsmith2024,
  author = {C. D. Arrowsmith and P. Simon and P. J. Bilbao and A. F. A. Bott and S. Burger and H. Chen and F. D. Cruz and T. Davenne and I. Efthymiopoulos and D. H. Froula and A. Goillot and J. T. Gudmundsson and D. Haberberger and J. W. D. Halliday and T. Hodge and B. T. Huffman and S. Iaquinta and F. Miniati and B. Reville and S. Sarkar and A. A. Schekochihin and L. O. Silva and R. Simpson and V. Stergiou and R. M. G. M. Trines and T. Vieu and N. Charitonidis and R. Bingham and G. Gregori},
  title = {Laboratory realization of relativistic pair-plasma beams},
  journal = {Nature Communications},
  volume = {15},
  pages = {5029},
  year = {2024},
  doi = {10.1038/s41467-024-49346-2}
}

@book{Bambi:2016lkv,
    editor = "Bambi, Cosimo",
    title = "{Astrophysics of Black Holes: From Fundamental Aspects to Latest Developments}",
    doi = "10.1007/978-3-662-52859-4",
    isbn = "978-3-662-52857-0, 978-3-662-52859-4",
    publisher = "Springer",
    address = "Berlin/Heidelberg",
    year = "2016"
}

@article{arons1979,
  title = {Some problems of pulsar physics or I'm madly in love with electricity},
  author = {Arons, Jonathan},
  journal = {Space Science Reviews},
  volume = {24},
  number = {4},
  pages = {437--510},
  year = {1979},
  publisher = {Springer}
}

@article{weidenspointner2008,
  title = {An asymmetric distribution of positrons in the Galactic disk revealed by $\gamma$-rays},
  author = {Weidenspointner, Georg and Skinner, Gerry and Jean, Pierre and Kn{\"o}dlseder, J{\"u}rgen and Von Ballmoos, Peter and Bignami, Giovanni and Diehl, Roland and Strong, Andrew W and Cordier, Bertrand and Schanne, St{\'e}phane and others},
  journal = {Nature},
  volume = {451},
  number = {7175},
  pages = {159--162},
  year = {2008},
  publisher = {Nature Publishing Group UK London}
}

@article{Meszaros2002,
   author = "Mészáros, P.",
   title = "Theories of Gamma-Ray Bursts", 
   journal= "Annual Review of Astronomy and Astrophysics",
   year = "2002",
   volume = "40",
   number = "Volume 40, 2002",
   pages = "137-169",
   doi = "https://doi.org/10.1146/annurev.astro.40.060401.093821",
   publisher = "Annual Reviews",
   issn = "1545-4282",
   type = "Journal Article",
  }

@article{hugenschmidt2012,
  title = {The NEPOMUC upgrade and advanced positron beam experiments},
  author = {Hugenschmidt, Christoph and Piochacz, Christian and Reiner, Markus and Schreckenbach, Klaus},
  journal = {New Journal of Physics},
  volume = {14},
  number = {5},
  pages = {055027},
  year = {2012},
  publisher = {IOP Publishing},
doi = {10.1088/1367-2630/14/5/055027},
}

@article{bernardini2004,
  title = {AdA: the first electron-positron collider},
  author = {Bernardini, Carlo},
  journal = {Physics in Perspective},
  volume = {6},
  pages = {156--183},
  year = {2004},
  publisher = {Springer},
  doi = {10.1007/s00016-003-0202-y},
}

@article{blumer2022,
  title={Positron accumulation in the {GBAR} experiment},
  author={Blumer, Philipp and Charlton, Michael and Chung, Moses and Clade, P and Comini, P and Crivelli, P and Dalkarov, O and Debu, P and Dodd, Liam and Douillet, A and others},
  journal={Nuclear Instruments and Methods in Physics Research Section A: Accelerators, Spectrometers, Detectors and Associated Equipment},
  volume={1040},
  pages={167263},
  year={2022},
  publisher={Elsevier},
  doi={https://doi.org/10.1016/j.nima.2022.167263},
}

@article{Chen2015,
  title = {Scaling the Yield of Laser-Driven Electron-Positron Jets to Laboratory Astrophysical Applications},
  author = {Chen, Hui and Fiuza, F. and Link, A. and Hazi, A. and Hill, M. and Hoarty, D. and James, S. and Kerr, S. and Meyerhofer, D. D. and Myatt, J. and Park, J. and Sentoku, Y. and Williams, G. J.},
  journal = {Phys. Rev. Lett.},
  volume = {114},
  issue = {21},
  pages = {215001},
  numpages = {5},
  year = {2015},
  month = {May},
  publisher = {American Physical Society},
  doi = {10.1103/PhysRevLett.114.215001},
}

@article{liang2015,
  title = {High e+/e- ratio dense pair creation with $10^{21} \; \mathrm{W/cm^2}$ laser irradiating solid targets},
  author = {Liang, E and Clarke, T and Henderson, A and Fu, W and Lo, W and Taylor, D and Chaguine, P and Zhou, S and Hua, Y and Cen, X and others},
  journal = {Scientific Reports},
  volume = {5},
  number = {1},
  pages = {13968},
  year = {2015},
  publisher = {Nature Publishing Group UK London},
  doi = {10.1038/srep13968},
}

@article{sarri2015,
  title = {Generation of neutral and high-density electron--positron pair plasmas in the laboratory},
  author = {Sarri, Gianluca and Poder, K and Cole, JM and Schumaker, W and Di Piazza, Antonino and Reville, Brian and Dzelzainis, T and Doria, D and Gizzi, LA and Grittani, G and others},
  journal = {Nature Communications},
  volume = {6},
  number = {1},
  pages = {6747},
  year = {2015},
  publisher = {Nature Publishing Group UK London},
  doi = {10.1038/ncomms7747},
}

@article{xu2016,
  title = {Ultrashort megaelectronvolt positron beam generation based on laser-accelerated electrons},
  author = {Xu, Tongjun and Shen, Baifei and Xu, Jiancai and Li, Shun and Yu, Yong and Li, Jinfeng and Lu, Xiaoming and Wang, Cheng and Wang, Xinliang and Liang, Xiaoyan and others},
  journal = {Physics of Plasmas},
  volume = {23},
  number = {3},
  year = {2016},
  publisher = {AIP Publishing},
  doi = {10.1063/1.4943280},
  pages = {033109}
}

@article{peebles2021,
  title = {Magnetically collimated relativistic charge-neutral electron--positron beams from high-power lasers},
  author = {Peebles, JL and Fiksel, G and Edwards, MR and von der Linden, J and Willingale, L and Mastrosimone, D and Chen, Hui},
  journal = {Physics of Plasmas},
  volume = {28},
  number = {7},
  year = {2021},
  publisher = {AIP Publishing},
  doi = {10.1063/5.0053557},
  pages = {074501}
}

@article{jiang2021,
  title = {Enhancing positron production using front surface target structures},
  author = {Jiang, Sheng and Link, Anthony and Canning, Dave and Fooks, JA and Kempler, PA and Kerr, Shaun and Kim, J and Krieger, Michael and Lewis, NS and Wallace, Russell and others},
  journal = {Applied Physics Letters},
  volume = {118},
  number = {9},
  year = {2021},
  publisher = {AIP Publishing},
  doi = {10.1063/5.0038222},
  pages = {094101},
}

@article{chen2023,
  title={Perspectives on relativistic electron--positron pair plasma experiments of astrophysical relevance using high-power lasers},
  author={Chen, Hui and Fiuza, Frederico},
  journal={Physics of Plasmas},
  volume={30},
  number={2},
  year={2023},
  publisher={AIP Publishing},
  doi={10.1063/5.0134819},
  pages={020601},
}

@article{Bethe1934,
    author = {Bethe, H.  and Heitler, W.  and Dirac, Paul Adrien Maurice },
    title = {On the stopping of fast particles and on the creation of positive electrons},
    journal = {Proceedings of the Royal Society of London. Series A},
    volume = {146},
    number = {856},
    pages = {83-112},
    year = {1934},
    doi = {10.1098/rspa.1934.0140},
}

@article{bohlen2014,
title = {The {FLUKA} Code: Developments and Challenges for High Energy and Medical Applications},
journal = {Nuclear Data Sheets},
volume = {120},
pages = {211-214},
year = {2014},
issn = {0090-3752},
doi = {https://doi.org/10.1016/j.nds.2014.07.049},
author = {T.T. Böhlen and F. Cerutti and M.P.W. Chin and A. Fassò and A. Ferrari and P.G. Ortega and A. Mairani and P.R. Sala and G. Smirnov and V. Vlachoudis},
}

@techreport{ferrari2005,
    author = "Ferrari, Alfredo and Sala, Paola R. and Fasso, Alberto and Ranft, Johannes",
    title = "{{FLUKA}: A multi-particle transport code (Program version 2005)}",
    institution = "CERN-2005-010, SLAC-R-773, INFN-TC-05-11",
    doi = "10.2172/877507",
    month = "10",
    year = "2005"
}

@article{fainberg1970,
  title = {Nonlinear theory of interaction between a monochromatic beam of relativistic electrons and a plasma},
  author = {Fainberg, Ya B and Shapiro, V and Shevchenko, V},
  journal = {Soviet Phys. JETP},
  volume = {30},
  pages = {528},
  year = {1970}
}

@article{ortner2022,
  title = {Numerically stable and computationally efficient expression for the magnetic field of a current loop},
  author = {Ortner, Michael and Leitner, Peter and Slanovc, Florian},
  journal = {Magnetism},
  volume = {3},
  number = {1},
  pages = {11--31},
  year = {2022},
  publisher = {MDPI},
  doi = {10.3390/magnetism3010002},
}

@software{magpylib,
  author       = {Ortner, Michael and others},
  title        = {Magpylib: A Python package for magnetic field computation,                            {https://magpylib.readthedocs.io/en/latest/}},
  year         = {2025},
  url          = {https://magpylib.readthedocs.io/en/latest/},
}

@incollection{gayde2024,
  author = {Jean-Christophe Gayde},
  title = {Survey and alignment of accelerators},
  booktitle = {Proceedings of the Joint Universities Accelerator School (JUAS II)},
  series = {CERN Yellow Reports: School Proceedings},
  volume = {3},
  number = {II},
  chapter = {11},
  location = {Geneva},
  year = {2024},
  doi = {10.23730/CYRSP-2024-003.1601},
  url = {https://e-publishing.cern.ch/index.php/CYRSP/article/view/1627},
  issn = {2519-805X},
  editor = {Elias M\'etral},
  pages = {1601-1635},
}

@article{aharonian2006,
  title={A low level of extragalactic background light as revealed by $\gamma$-rays from blazars},
  author={Aharonian, F and Akhperjanian, AG and Bazer-Bachi, AR and Beilicke, M and Benbow, W and Berge, D and Bernl{\"o}hr, K and Boisson, C and Bolz, O and Borrel, V and others},
  journal={Nature},
  volume={440},
  number={7087},
  pages={1018--1021},
  year={2006},
  publisher={Nature Publishing Group UK London},
  doi={10.1038/nature04680},
}

@article{broderick2012,
  title={The cosmological impact of luminous {TeV} blazars. {I}. {I}mplications of plasma instabilities for the intergalactic magnetic field and extragalactic gamma-ray background},
  author={Broderick, Avery E and Chang, Philip and Pfrommer, Christoph},
  journal={The Astrophysical Journal},
  volume={752},
  number={1},
  pages={22},
  year={2012},
  publisher={IOP Publishing},
  doi={10.1088/0004-637X/752/1/22},
}

@article{schlickeiser2012,
  title={Plasma effects on fast pair beams in cosmic voids},
  author={Schlickeiser, R and Ibscher, D and Supsar, M},
  journal={The Astrophysical Journal},
  volume={758},
  number={2},
  pages={102},
  year={2012},
  publisher={IOP Publishing},
  doi={10.1088/0004-637X/758/2/102},
}

@article{neronov2010,
  title={Evidence for strong extragalactic magnetic fields from Fermi observations of {TeV} blazars},
  author={Neronov, Andrii and Vovk, Ievgen},
  journal={Science},
  volume={328},
  number={5974},
  pages={73--75},
  year={2010},
  publisher={American Association for the Advancement of Science},
  doi={10.1126/science.1184192},
}

@InProceedings{fonseca2002,
author="Fonseca, R. A.
and Silva, L. O.
and Tsung, F. S.
and Decyk, V. K.
and Lu, W.
and Ren, C.
and Mori, W. B.
and Deng, S.
and Lee, S.
and Katsouleas, T.
and Adam, J. C.",
editor="Sloot, Peter M. A.
and Hoekstra, Alfons G.
and Tan, C. J. Kenneth
and Dongarra, Jack J.",
title="OSIRIS: A Three-Dimensional, Fully Relativistic Particle in Cell Code for Modeling Plasma Based Accelerators",
booktitle="Computational Science --- ICCS 2002",
year="2002",
publisher="Springer",
address="Berlin, Heidelberg",
pages="342--351",
isbn="978-3-540-47789-1",
doi={10.1007/3-540-47789-6_36},
}

@inproceedings{efthymiopoulos2011,
author = {I. Efthymiopoulos and C. Hessler and H. Gaillard and D. Grenier and M. Meddahi and P. Trilhe and A. Pardons and
C. Theis and N. Charitonidis and S. Evrard and H. Vincke and M. Lazzaroni},
editor = {Christine Petit-Jean-Genaz},
title = {{HiRadMat}: {A} New Irradiation Facility for Material Testing at {CERN}},
booktitle = {2nd International Particle Accelerator Conference, San Sebastian, Spain, 4 -- 9. September 2011},
pages = {1665--1667},
year = {2011},
publisher = { },
address = {},
url={https://jacow.org/IPAC2011/papers/tups058.pdf},
}

@article{slezak2016,
  title={Temperature-wavelength dependence of terbium gallium garnet ceramics Verdet constant},
  author={Slez{\'a}k, Ond{\v{r}}ej and Yasuhara, Ryo and Lucianetti, Antonio and Mocek, Tom{\'a}{\v{s}}},
  journal={Optical Materials Express},
  volume={6},
  number={11},
  pages={3683--3691},
  year={2016},
  publisher={Optical Society of America}, 
  doi={10.1364/OME.6.003683}
}

@article{arrowsmith2025,
author = {Charles D. Arrowsmith and Francesco Miniati and Pablo J. Bilbao and Pascal Simon and Archie F. A. Bott and Stephane Burger and Hui Chen and Filipe D. Cruz and Tristan Davenne and Anthony Dyson and Ilias Efthymiopoulos and Dustin H. Froula and Alice Goillot and Jon T. Gudmundsson and Dan Haberberger and Jack W. D. Halliday and Tom Hodge and Brian T. Huffman and Sam Iaquinta and G. Marshall and Brian Reville and Subir Sarkar and Alexander A. Schekochihin and Luis O. Silva and Raspberry Simpson and Vasiliki Stergiou and Raoul M. G. M. Trines and Thibault Vieu and Nikolaos Charitonidis and Robert Bingham and Gianluca Gregori},
title = {Suppression of pair beam instabilities in a laboratory analogue of blazar pair cascades},
journal = {Proceedings of the National Academy of Sciences},
year = 2025,
volume = 122,
number = 45,
pages = {e2513365122},
doi = {10.1073/pnas.2513365122}
}

@article{silva2002,
  title={On the role of the purely transverse Weibel instability in fast ignitor scenarios},
  author={Silva, Lu{\i}́s O and Fonseca, Ricardo A and Tonge, John W and Mori, Warren B and Dawson, John M},
  journal={Physics of Plasmas},
  volume={9},
  number={6},
  pages={2458--2461},
  year={2002},
  publisher={American Institute of Physics},
  doi={10.1063/1.1476004},
}

@article{bret2010,
  title={Multidimensional electron beam-plasma instabilities in the relativistic regime},
  journal={Physics of Plasmas},
  volume={17},
  number={12},
  year={2010},
  publisher={AIP Publishing},
  doi={10.1063/1.3514586},
}

\end{document}